# Qualified Trust, not Surveillance, is the Basis of a Stable Society

By Dirk Helbing, ETH Zurich, Clausiusstrasse 50, 8092 Zurich, Switzerland

**Peaceful citizens and hard-working taxpayers are under government surveillance. Confidential communication of journalists is intercepted. Civilians are killed by drones, without a chance to prove their innocence. How could it come that far? And what are the alternatives?**

Since September 11, 2001, freedom rights have been restricted in most democracies step by step. Each terrorist threat has delivered new reasons to extend the security infrastructure, which is eventually reaching Orwellian dimensions [1]. Through its individual configuration, every computer has an almost unique fingerprint, allowing one to record our use of the Web. Privacy is gone. Over the past years, up to 1500 variables about half a billion citizens in the industrial world have been recorded. Google and facebook know us better than our friends and families.

Nevertheless, governments have failed so far to gain control of terrorism, drug traffic, cybercrime and tax evasion. Would an omniscient state be able to change this and create a new social order? [3] It seems at least to be the dream of secret services and security agencies.

Ira "Gus" Hunt, the CIA Chief Technology Officer, recently said [2]:

> *"You're already a walking sensor platform… You are aware of the fact that somebody can know where you are at all times because you carry a mobile device, even if that mobile device is turned off. You know this, I hope? Yes? Well, you should… Since you can't connect dots you don't have, it drives us into a mode of, we fundamentally try to collect everything and hang on to it forever… It is really very nearly within our grasp to be able to compute on all human generated information."*

Unfortunately, connecting the dots often does not work. As complex systems experts point out, such "linear thinking" can be totally misleading. It's the reason why we often want to do the right things, but take the wrong decisions.

I agree that our world has destabilized. However, this is not a result of external threats, but of system-immanent feedback effects. The increasing interdependencies, connectivity and complexity of our world and further trends are causing this [4]. However, trying to centrally control this complexity is destined to fail. We must rather learn to embrace the potential of complexity. This requires a step towards decentralized self-regulatory approaches. Many of us believe in Adam Smiths "invisible hand", according to which the best societal and economic outcome is reached, if everybody is just doing what is best for himself or herself. However, this principle is known to produce "market failures", "financial meltdowns", and other "tragedies of the commons"

(such as environmental degradation). The classical approach is to try to "fix" these problems by top-down regulation of a powerful state.

However, self-regulation based on decentralized rules *can* be learned. This has been demonstrated for modern traffic control concepts, but it's equally relevant for smart energy grids, and will be even more important for the financial system. The latter, for example, needs built-in breaking points similar to the fuses in our electrical network at home, and it requires additional control parameters to equilibrate.

There *is* an alternative to uncoordinated bottom-up organization and too much top-down regulation -- a better one: the "economy 2.0". Doing the step towards a self-regulating, participatory market society can unleash the unused potential of complexity and diversity, which we are currently trying to fight [5]. This step can boost our societies and economies as much as the transition from centrally regulated societies to market societies inspired by Adam Smith. But after 300 years, it's now time for a new paradigm. Societies based on surveillance and punishment are not long-term sustainable. When controlled, people get angry, and the economy never thrives. Qualified trust is a better basis of resilient societies. But how to build it? Reputation systems are now spreading all over the web. If properly designed, they could be the basis of a self-regulating societal and market architecture. Further success principles of decentralized self-regulating systems can be learned from ecological and immune systems. They could be also the basis of a trustable Web, which would protect itself from harmful actions and contain cybercrime.

Rather than in surveillance technology, government should invest their money in the creation of self-regulating architectures. It will be crucial for a successful transition to a new era -- the era of information societies. If we take the right decisions, the 21st century can be an age of creativity, prosperity and participation. But if we take the wrong decisions, we will end in economic and democratic depression. It's our choice.